\documentclass[twocolumn]{aastex62}
\usepackage{amsmath,amssymb,bm}

\shorttitle{New Graph Diagnostics : $\alpha, \tau_\Delta, n_{s\ge5}$}
\shortauthors{Hong et al.}

\begin{document}


\title{Constraining Cosmology with Big Data Statistics of Cosmological Graphs} 


\author{Sungryong Hong} 
\affiliation{School of Physics, Korea Institute for Advanced Study,
85 Hoegiro, Dongdaemun-gu, Seoul 02455, Korea}

\author{Donghui Jeong} 
\affiliation{Department of Astronomy and Astrophysics and Institute for Gravitation and the Cosmos, The Pennsylvania State University,
University Park, PA 16802, USA}

\author{Ho Seong Hwang}
\affiliation{Quantum Universe Center, Korea Institute for Advanced Study,
85 Hoegiro, Dongdaemun-gu, Seoul 02455, Korea}
\affiliation{Korea Astronomy and Space Science Institute,
776 Daedeokdae-ro, Yuseong-gu, Daejeon 34055, Korea}

\author{Juhan Kim} 
\affiliation{Center for Advanced Computation, Korea Institute for Advanced Study, 
85 Hoegiro, Dongdaemun-gu, Seoul 02455, Republic of Korea}

\author{Sungwook E. Hong}
\affiliation{Korea Astronomy and Space Science Institute,
776 Daedeokdae-ro, Yuseong-gu, Daejeon 34055, Korea}
\affiliation{Natural Science Research Institute, University of Seoul,
163 Seoulsiripdaero, Dongdaemun-gu, Seoul 02504, Republic of Korea}

\author{Changbom Park} 
\affiliation{School of Physics, Korea Institute for Advanced Study,
85 Hoegiro, Dongdaemun-gu, Seoul 02455, Korea}

\author{Arjun Dey} 
\affiliation{National Optical Astronomy Observatory, 950 N. Cherry Ave., 
Tucson, AZ 85719, USA}

\author{Milos Milosavljevic} 
\affiliation{Department of Astronomy, The University of Texas at Austin, 
Austin, TX 78712, USA}

\author{Karl Gebhardt} 
\affiliation{Department of Astronomy, The University of Texas at Austin, 
Austin, TX 78712, USA}

\author{Kyoung-Soo Lee} 
\affiliation{Department of Physics and Astronomy, Purdue University, 
525 Northwestern Avenue, West Lafayette, IN 47907, USA}

\begin{abstract}
By utilizing large-scale graph analytic tools implemented in the modern Big Data platform, \textsc{Apache Spark}, 
we investigate the topological structure of gravitational clustering in five different universes produced by cosmological $N$-body simulations 
with varying parameters: 
(1) a \replaced{standard universe}{WMAP 5-year compatible $\Lambda$CDM cosmology}, (2) two different dark energy equation of state variants, and (3) two different cosmic matter density variants.
For the Big Data calculations, we use a custom build of stand-alone Spark/Hadoop cluster at Korea Institute for Advanced Study (KIAS) 
and Dataproc Compute Engine in Google Cloud Platform (GCP) with the sample size ranging from 7 millions to 200 millions.
We find that among the many possible graph-topological measures, three
simple ones: (1) \added{the average of number of neighbors (the so-called average vertex degree)} $\alpha$, (2) \added{closed-to-connected triple fraction (the so-called transitivity)} $\tau_\Delta$, and (3) \added{the cumulative number density $n_{s\ge5}$ of subcomponents with connected component size $s \ge 5$}, 
can effectively discriminate among the five model universes. Since these graph-topological measures are in direct relation with the usual $n$-points correlation functions of the cosmic density field, graph-topological statistics powered by Big Data computational infrastructure opens a new, intuitive, and computationally efficient window into the dark Universe.
\end{abstract}



\section{Introduction}
 
The evolution of the Universe has imprinted various unique patterns of spatial organization in the cosmic matter distribution.  Patterns that have appeared and disappeared across cosmic epochs are accessible to us in the Big Data that is being acquired with astronomical surveys. 
For understanding the genesis of the Universe and its evolution to the present epoch it is important to extract all the  information that is latent in survey data. Subtle diagnostics of spatial organization have the promise to break formidable degeneracies in our picture of the quantum Universe and the nature of gravity. 

During the last two decades, studies of the angular anisotropy
of the cosmic microwave background (CMB) have  provided support for the so-called $\Lambda$CDM cosmological model \citep[e.g.,][]{WMAP5, Planck}
and have elevated cosmology to an unprecedented level of precision. 
The baryon acoustic feature, also known as baryon acoustic oscillations \citep[BAO; e.g.,][]{eisenstein, eisenstein05,HETDEX, BOSS, DESI, eBOSS}, 
has been shown to be an effective ``standard ruler'' that captures geometric information that is indicative of the universal expansion rate. Numerous galaxy surveys are being performed as well as planned to measure the BAO feature by mapping out the matter distribution of the Universe on large-scales.

The successful measurements of the CMB angular power spectrum and BAO feature show how useful two-point statistics have been for quantifying the geometry of the universe and the evolution of cosmic structure. 
The more challenging higher-order statistical measurements, such as of the three- and four-point correlation functions (or bi- and tri-spectra of density fluctuations in Fourier space),
can provide powerful further constraints that can shed light on a hypothetical non-Gaussianity of primordial quantum fluctuations \citep[e.g.,][]{takahashi,nongplanck}. 
The pursuit of primordial non-Gaussianity is just one example how $n$-point statistics provides a unique window into the fundamental physical substrate of the observed Universe.

Along with the successful $n$-point correlation functions, 
various topological measures have been introduced, 
such as Betti numbers, Minkowski functionals, and genus statistics
\citep[][]{gott87,eriksen04,parkbetti,weygaert13,pranav17}.
To identify specific topological structures such as cosmic filaments and voids,
many techniques have been attempted, such as for example wavelets, minimum-spanning trees, 
Morse theory, watershed transforms, and smoothed Hessians \citep[e.g.,][]{barrow,seth,martinez,aragon,colberg,sousbie,bond,cautun}. 
While these topological methods can provide valuable insight, they are generally ad hoc and not (yet) justified within a principled and physically rigorous framework, the kind of framework that justifies the successful $n$-point statistical approaches.

\begin{deluxetable*}{c|cc|ccc}[t!]
\tablecaption{Hardware Configurations for the Spark Clusters\tablenotemark{$\dagger$} \label{tab:hwspec}}
\tablecolumns{6}
\tablenum{1}
\tablewidth{0pt}
\tablehead{
\colhead{} & \multicolumn{2}{c}{Driver Node} & \multicolumn{3}{c}{Worker Node} \\
\colhead{Cluster Name} & \colhead{vCPUs\tablenotemark{$\dagger$}} &
\colhead{Memory}  & \colhead{vCPUs\tablenotemark{$\dagger$}} & \colhead{Memory} & \colhead{$n$Workers\tablenotemark{$\dagger$}}
}
\startdata
KIAS Standalone\tablenotemark{a} & 4 & 32GB  & 16 & 52GB & 3 \\
Google Cloud Dataproc\tablenotemark{b} & 16 & 104GB & 32 & 208GB  & 5\\
\enddata
\tablenotetext{\dagger}{Generally, a Spark cluster is composed of one driver node and multiple worker nodes. 
vCPUs represents the number of {\it logical} cores (e.g., {\it hyperthreading}) for each node and $n$Workers is the number of worker nodes in each cluster.}
\tablenotetext{a}{The KIAS standalone cluster is custom-built by adding three Linux worker nodes 
to a Mac OS X driver node.}  
\tablenotetext{b}{Cloud Dataproc is a cloud service for running cloud-native Apache Hadoop/Spark clusters in Google Cloud Platform. 
Since we are allowed to create and resize Spark clusters within the available quota of 192 vCPUs and 2048 GB memory, 
Google Dataproc can compensate for the limited capacity of our standalone cluster.}
\end{deluxetable*}

As a new way to quantify the elusive topological structure of the Universe, 
here we apply graph theory (or, {\it network science}) to cosmological datasets \citep{hong15,hong16,hong19}. 
The basic idea is to associate galaxies with the vertices of a graph and to connect nearby galaxies with graph edges. Then we compute graph-theoretic statistical measures of the cosmic matter distribution as traced by galaxies.
We have previously proposed and tested various graph-theoretic  topological diagnostic indicators on cosmological datasets, but our attempts to-date were limited to insufficient datasets, ones that were small enough to fit in the memory of workstations.  Here, we embrace bleeding-edge technology to overcome this restriction and analyze datasets large enough to extract cosmologically-discriminative statistical indicators.

In this paper, by utilizing the modern Big Data platform, \textsc{Apache Spark} \citep{Zaharia:EECS-2014-12, sparkphysics}, 
we investigate the topological structure of five different universes, all generated with cosmological $N$-body simulations 
with various input parameters but seeded with same realization of a Gaussian random field. The galaxy sample size extracted from the simulations ranged from 7 million to 200 million galaxies.
To calculate graph statistics of these Big Data samples, 
we built our own stand-alone Spark/Hadoop cluster at the Korea Institute for Advanced Study (KIAS)
and also used the commercial cloud cluster, the Cloud Dataproc service within the Google Cloud Platform (GCP), 
for some of the calculations that required more computation resources than what the KIAS stand-alone cluster could provide. 
We summarize the hardware specifications of these clusters in Table~\ref{tab:hwspec}. 

This paper is organized as follows. 
In Section 2, we describe our $N$-body simulations, which we name {\it Multiverse}, and how we generated graphs from the simulation data.  
In Section 3, we present a mathematical formulation of the graph statistical methods and  
in Section 4 we apply the methods to our datasets and propose a diagnostic scheme that discriminates between the five different universes. 
Finally, in Section 5, we summarize our results and list our conclusions.
We interchangeably use the terminology of graph theory and network science, such as {\it vertex} vs.\ {\it node}, 
{\it edge} vs.\ {\it link}, and {\it graph} vs.\ {\it network}.

\section{DATA}\label{sec:data}

\subsection{Multiverse Simulations}

The Multiverse Simulations are a set of cosmological pure $N$-body simulations designed to study the effect of cosmological parameters 
on the formation of large-scale structures (LSS) in various universe models as traced by galaxies (Kim et al. in prep.).
The fiducial simulation is based on the concordance $\rm \Lambda$CDM model with $H_0=100\,h\,\text{km}\,\text{s}^{-1}\,\text{Mpc}^{-1}$ where 
$h=0.72$, $\Omega_{\rm m}=0.26$, $\Omega_\Lambda=0.74$, $\Omega_{\rm b}=0.044$, $w=-1$ and
$b_8=1.2$ (hereafter, we refer to this fiducial universe as ``standard universe'' denoted by STD). 
Here, $w$ is the pressure-to-energy density ratio that parametrizes the equation of state of the dark energy. 
The shape of linear power spectrum was obtained from the CAMB code and its power spectral amplitude 
is tuned to make the density fluctuations satisfy the relation
$\sigma_8 \equiv 1/b_8$. Here, $\sigma_8$ is the standard deviation of the density field when 
smoothed with a top-hat spherical kernel with radius $R_{\rm tophat}= 8 ~h^{-1}\,\text{Mpc}$ at $z=0$.
We placed the simulation particles at grid points as pre-initial conditions 
and perturbed them using the second-order linear perturbation method.  
The gravitational evolution of particles was performed with
the GOTPM code \citep{2004NewA....9..111D} that solves 
the Poisson equation with the Fast-Fourier-Transforms (FFT) and corrects the short-range force with the Barnes-Hut tree method.

For the non-standard-$\Lambda$CDM simulations, we adopt four variant models different from our fiducial $\Lambda$CDM in a single parameter:
\begin{itemize}
    \item DM1: $\Omega_{\rm m}=0.31$,
    \item DM2: $\Omega_{\rm m}=0.21$,
    \item DE1: $w =-0.5$,
    \item DE2: $w =-1.5$.
\end{itemize}
The same random number sequence is applied to generating initial conditions, which may eliminate the cosmic variances between simulated models.
Therefore, it would be possible to study the pure cosmological effects on  structure and galaxy formation by directly comparing the distributions of cosmic objects.
The number of particles in each simulation is $N_{\rm p}=2048^3$. We integrate the gravitational evolution of the models starting redshift of $z_{\rm init}=99$ to the final epoch $z=0$
with 1980 steps.
The simulation box size is $L_{\rm box} = 1024 ~h^{-1}\,\text{Mpc}$ in the comoving scale.

\begin{figure*}[t]
\centering
\includegraphics[height=4.8 in]{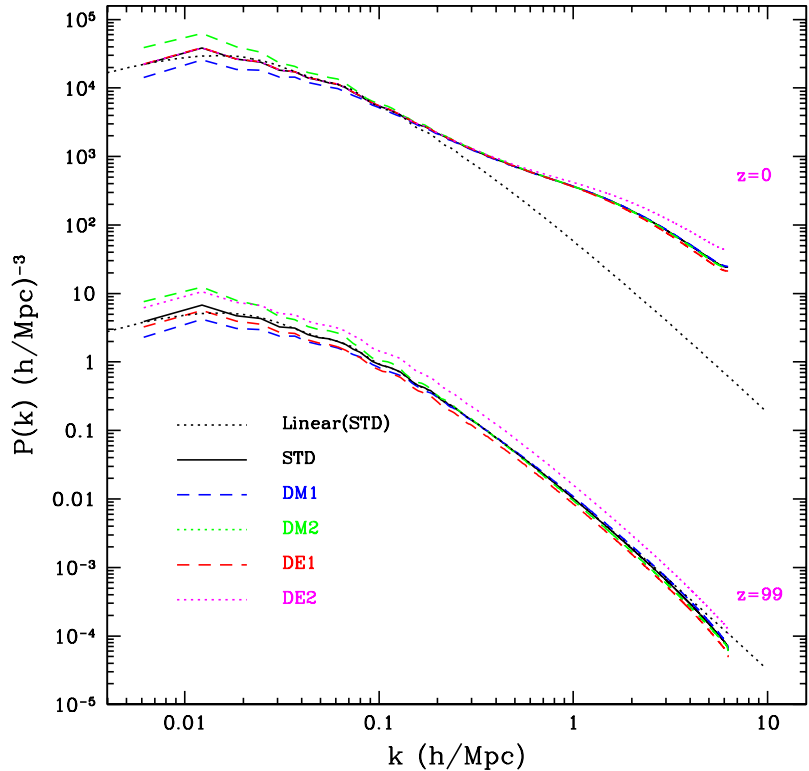}
\caption{Simulated matter power spectra of the Multiverse Simulations 
at $z_{\rm init}=99$ and $z=0$.
The dotted lines are the linear power spectrum of the standard universe, STD.
}\label{fig:power}
\end{figure*} 
Figure \ref{fig:power} shows the simulated mass power spectra ({\it{colored solid lines}})
of the Multiverse simulations compared
to the linear expectation of the $\rm \Lambda$CDM model ({\it dotted lines}).
At $z=0$, the small-scale power spectrum of DE2 has a relatively higher amplitude than that of the other simulations. This difference is mainly due to the higher power amplitude
of DE2 at the starting redshift that makes the small-scale perturbation enter the
nonlinear regime earlier.

\begin{figure*}[t]
\centering
\includegraphics[height=3.0 in]{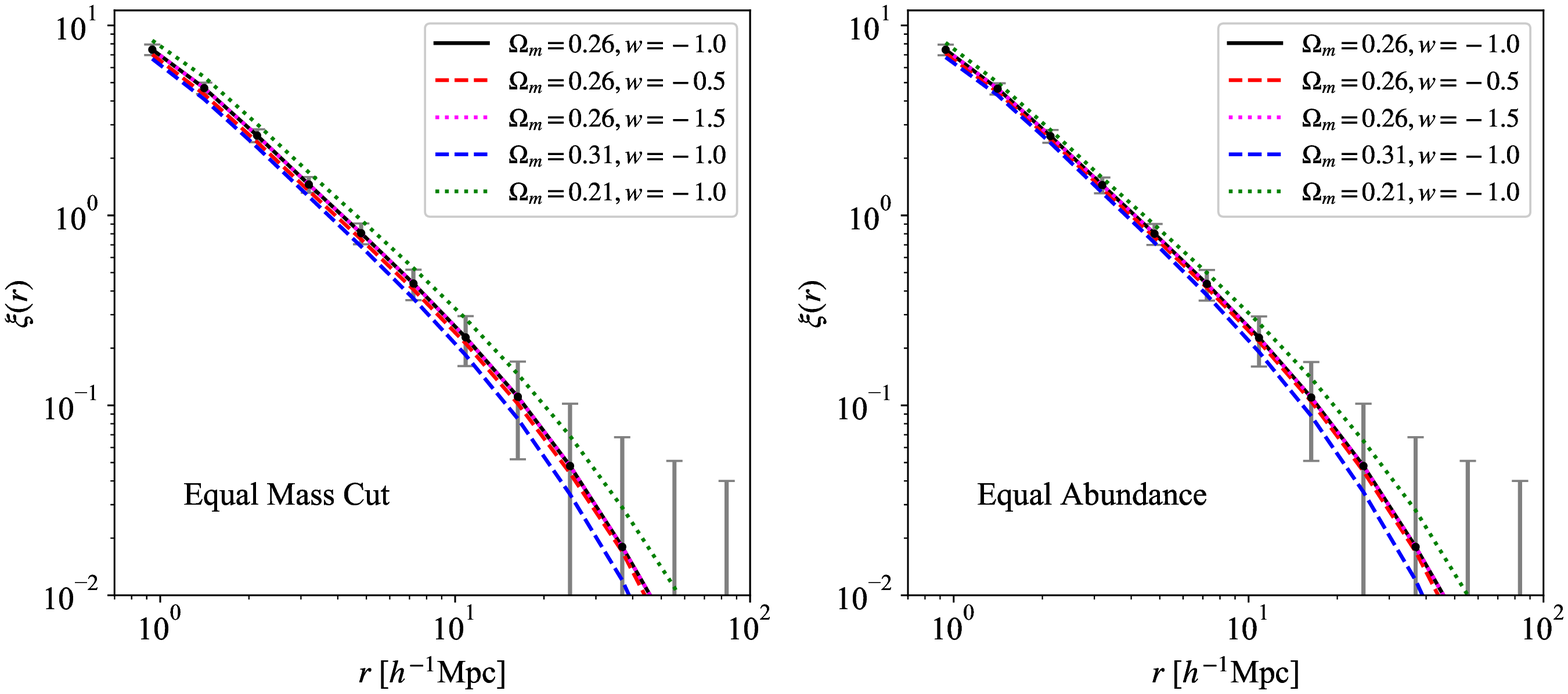}
\caption{The two-point correlation functions for equal mass cut sample (left), $M_{\rm cut} = 5 \times 10^{11} h^{-1}$ M$_\odot$, 
and equal abundance sample (right), $N_{h} = 7,086,717$. The grey error bars represent 
the bootstrap resampling errors for STD. The other Multiverses show similar bootstrap errors, 
skipped in the panels due to the redundancy.
}\label{fig:twopoint}
\end{figure*} 
For generating halo catalogs, we extract virialized halos with the minimum mass of
$M_{\rm min}=2.7\times 10^{11} (\Omega_{\rm m}/0.26) ~h^{-1}{\rm M_\odot}$, which corresponds to a minimum of
30 particles. We have used the standard Friends-of-Friends \added{(FoF)} method 
with linking length $l_{\rm FoF}=0.2\times l_{\rm mean}$ 
where $l_{\rm mean}$ is the average distance between particles.

From the halo catalogs, we select two kinds of samples with  
(1) equal mass cut, $M_{\rm cut} = 5 \times 10^{11} h^{-1}$ M$_\odot$, and 
(2) equal abundance cut, $N_{\rm h} = 7,086,717$, corresponding to a comoving density of $n_{\rm h} = 6.6\times10^{-3}  \big[ h^{-1}$Mpc$\big]^{-3}$,  
as summarized in Table~\ref{tab:sample}. 
Figure~\ref{fig:twopoint} shows the two-point correlation function for each halo selection criterion. 
The grey error bars represent the conventional bootstrap resampling errors  for STD. 

For graph measurements, any reshuffling by resampling  can affect the graph connectivity.
Therefore, instead of resampling to measure cosmic variances of graph statistics, 
we use a halo catalog from Horizon Run 4 \citep[][hereafter, \replaced{Horizon Run or HR}{STD-HR}]{hr4}, which has the same cosmological 
parameters as STD, but a much larger volume of ($3,150$ $h^{-1}$ Mpc)$^3$. 
Hence, at least for STD, we can measure the comic variance of graph statistics directly 
by subsampling the \replaced{Horizon Run}{STD-HR} catalog. 
Thanks to \textsc{Apache Spark} we can easily handle this Big Data catalog that is composed of 206 millions halos.

\begin{deluxetable*}{cc|rr|rr}[t!]
\tablecaption{Sample Selections \label{tab:sample}}
\tablecolumns{6}
\tablenum{2}
\tablewidth{0pt}
\tablehead{
\multicolumn{2}{c}{Multiverses} & \multicolumn{2}{c}{Equal Mass Cut Sample} & \multicolumn{2}{c}{Equal Abundance Sample\tablenotemark{$a$}} \\
\colhead{Name} & \colhead{Cosmological Parameters} & \colhead{$N_{h}$} &
\colhead{$M_{\rm cut}$($h^{-1}$ M$_\odot$)} & \colhead{$N_{h}$} & \colhead{$M_{min}$($h^{-1} $M$_\odot$)} 
}
\startdata
STD & $\Omega_{\rm m} = 0.26, w = -1.0$ & 7,086,717 & $5.00\times 10^{11}$  & 7,086,717 & $5.05\times 10^{11}$ \\
\hline
DE1 & $\Omega_{\rm m} = 0.26, w = -0.5$ & 7,806,135 & $5.00\times 10^{11}$ & 7,086,717 & $5.59\times 10^{11}$\\
DE2 & $\Omega_{\rm m} = 0.26, w = -1.5$ & 6,886,870 & $5.00\times 10^{11}$ & 7,086,717 & $4.87\times 10^{11}$\\
\hline
DM1 & $\Omega_{\rm m} = 0.31, w = -1.0$ & 8,595,923 & $5.00\times 10^{11}$ & 7,086,717 & $6.24\times 10^{11}$\\
DM2 & $\Omega_{\rm m} = 0.21, w = -1.0$ & 5,579,491 & $5.00\times 10^{11}$ & 7,086,717 & $3.86\times 10^{11}$\\
\hline
\hline
STD-HR & Horizon Run \added{4}\tablenotemark{$\dagger$} & 206,140,716 & $5.00\times 10^{11}$  & 206,140,716 & $5.05\times 10^{11}$ \\
\enddata
\tablenotetext{a}{The comoving density for $N_{h} = 7,086,717$ is $n_{\rm h} = 6.6\times10^{-3}  \big[ h^{-1}$Mpc$\big]^{-3}$ 
and its average distance $\langle r\rangle \sim n_{\rm h}^{-\frac{1}{3}} = 5.3 h^{-1}$Mpc. }
\tablenotetext{\dagger}{The cosmological parameters of Horizon Run \added{4}, $\Omega_{\rm m} = 0.26$ and $w = -1.0$, are the same 
with the standard universe, STD, in the Multiverse simulations. The difference is 
the Horizon \added{Run 4}'s huge volume\deleted{ size}, ($3,150$ $h^{-1}$ Mpc)$^3$, which is 29 times larger than 
the \deleted{standard universe in the}Multiverse suite. }
\end{deluxetable*}

\subsection{Generating Halo Networks}

To build a network from each halo distribution, 
we use the conventional FoF recipe \citep{huchra,hong15,hong16,hong19}. 
For a given linking length $l$, the {\it adjacency matrix} of the FoF recipe can be written as, 
\begin{equation}\label{eq:adj}
A_{ij} = \left\{ \begin{array}{ll}
	1 & \textrm{   if   } r_{ij} \le l,  \\
	0 & \textrm{   otherwise, } 
	\end{array} \right.
\end{equation}
where $r_{ij}$ is the distance between the two vertices (i.e., galaxies), $i$ and $j$. 
This binary matrix is essential in graph analysis as it quantifies network connectivity. 
Interested readers can consult \cite{barr}, \cite{newmanr}, \cite{dorogr}, and \cite{barthr} 
for further information.

 \section{Statistics of Graph Configurations}\label{sec:3}
 
In this section, we present basic graph quantities and their definitions used in network science \citep{dall,barthr}. 
Then, we show how each graph quantity is related to $n$-point correlation functions. 
The details of mathematical derivations can be found in a separate paper (Jeong et al. 2019, in prep.). 

\subsection{Basic Quantities}

First, we define two basic quantities, 
 \begin{eqnarray}
\alpha & \equiv & \frac{2K}{N},\label{eq:basic1} \\
p & \equiv & \frac{2K}{N (N-1)},\label{eq:basic2}
\end{eqnarray}
where $N$ is the total number of vertices and $K$ the total number of edges. 
We define $degree$ as the number of neighbors for each vertex. 
Then, $\alpha$ means the {\it average} of all degrees for the network; 
generally referred to as {\it average degree} in network science.   
$p$ is the fraction of real connected edges out of the total pair-wise combinations, $N(N-1)/2$; 
referred to as {\it edge density}.  
Finally, $\alpha$ and $p$ satisfy this trivial equality, 
\begin{eqnarray}\label{eq:basictwo}
\alpha & = & p (N-1). 
\end{eqnarray}

\subsubsection{Ensemble Average and Random Poisson Graph}

If we can define an ensemble of graphs, we can derive many graph statistics 
from probability distribution functions based on ensemble averages. 
Let us assume that we have a graph ensemble, denoted by $G_{\alpha,p}$ for given $\alpha$ and $p$. 
The average degree, $\alpha$, now can be written using a degree distribution, $p_k$, as  
\begin{eqnarray}
\alpha & = & \sum_{k=0}^{\infty} k \times p_k,  
\end{eqnarray}
where $k$ is a degree and $p_k$ a probability density for given $k$ with the normalization of $\sum_{k=0}^{\infty} p_k = 1$. 
If we randomly connect two vertices using the probability, $p$, in Equation~\ref{eq:basic2} 
(i.e., generating random graphs), 
the degree distribution of this ensemble is Poissonian, 
\begin{eqnarray}
p_k & \simeq & \frac{\alpha^k e^{-\alpha}}{k!}, 
\end{eqnarray}
in the limit of large $N$. 
To discern these random graphs from random geometric graphs in the following section, 
we refer to this kind as Random Poisson Graph (RPG).

\subsubsection{Geometric Graphs and Correlation Functions}

Now, we consider a graph embedded in a metric space; specifically, in this paper, $d$\added{-}dimensional Euclidean space. 
Since RPG described in the previous section has no geometric restriction, it can be described by only two parameters, 
$N$ and $K$; or, corresponding $\alpha$ and $p$.

For geometric graphs, we have additional quantities\replaced{,}{:} (1) spatial dimension, $d$,  (2) total system volume, $V$, 
and (3) linking length for connections, $l$, along with $N$ and $K$. 
Based on these parameters, determining geometric graphs, we define three basic quantities, 
the spatial number density, $\bar{n}$, 
excluded volume\footnote{The terminology of {\it excluded volume} is adopted from continuum percolation theory, 
which defines the connections in FoF networks.}, 
$V_l$, and fraction of excluded volume, $q$, 
\begin{eqnarray}
\bar{n} & \equiv & \frac{N}{V}, \\
V_l & \equiv & \frac{\pi^{d/2} l^d}{\Gamma \big(\frac{d+2}{2}\big)},\label{eq:exvol}  \\
q & \equiv & \frac{V_l}{V},
\end{eqnarray}
where $\Gamma(x)$ is the gamma function. 
 Then, for $d=3$, $\alpha$ and $p$ can be derived as, 
\begin{eqnarray}
\alpha & = & \bar{n} \int_{V_l} d^3r[1 + \xi(r)], \label{eq:10} \\ 
p & \simeq & \frac{1}{V} \int_{V_l} d^3r[1 + \xi(r)] \label{eq:11},  
\end{eqnarray}
using 2-point correlation function, $\xi(r)$ (Jeong et al., in prep.).

Unlike the simple derivations of $\alpha$ and $p$, 
the degree distribution, $p_k$, is inevitably 
complex determined by all orders of correlation functions, 
\begin{eqnarray}
p_k & \sim & \mathcal{F}( \{C_{k=1,2,\cdots} \}) \label{eq:12}, 
\end{eqnarray}
where $C_k$ represents $k$-point correlation function.  
\added{On the other hand,}
since random geometric graphs (RGGs) have null correlation functions, 
Equation~\ref{eq:10}, \ref{eq:11}, and~\ref{eq:12} for RGGs are as simple as,   
\begin{eqnarray}
\alpha & = & \bar{n} V_l, \\
p & \simeq & q, \\
p_k & \simeq & \frac{\alpha^k e^{-\alpha}}{k!}.  
\end{eqnarray}
Hence, any deviations of cosmological networks from these RGGs are caused 
by the non-zero correlation functions of cosmic datasets.

\begin{figure*}[t]
\centering
\includegraphics[height=2.4 in]{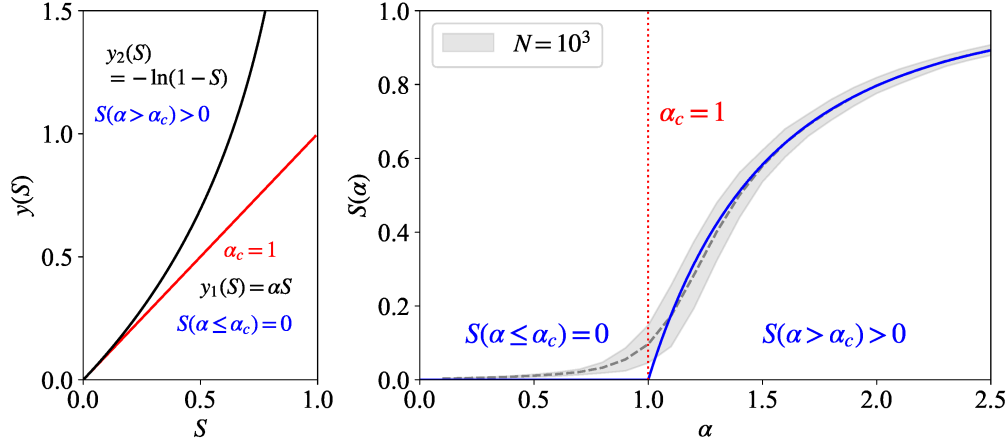}
\caption{The percolation threshold, $\alpha_c$, and giant component fraction, $S$, for RPGs. 
The left panel demonstrates which $\alpha$ results in the non-zero solution of 
giant component fraction, $S$. 
The right panel summarizes the analytic solutions in the plot of $S(\alpha)$ vs. $\alpha$. 
The grey dashed line represents the mean values of simulated results for RPGs with $N=10^3$ vertices, and 
the grey shaded area the $\pm 1 \sigma$ variations. 
In the Big Data regime, i.e., $N \to \infty$, the simulated giant component fractions 
converge to the theoretical line. 
}\label{fig:randpercol}
\end{figure*}

\subsection{Giant Component and Percolation Threshold}

The {\it giant component} is the largest connected subgraph in a network. 
The fraction, $S$, of vertices belonging to the giant component 
can be written using a generating function, $G_0(x)$, as 
\begin{eqnarray}
S & = & 1 - G_0(u), \label{eq:gone}\\
G_0(x) & = & \sum_{k=0}^{\infty} p_k x^k, \label{eq:gtwo}
\end{eqnarray}
where $u$ is the probability of a vertex, not belonging to the giant component, which satisfies a self-consistent equation,
\begin{eqnarray}
u & = & G_1(u), \label{eq:gthree}
\end{eqnarray}
where $G_1(x) = G'_0(x)/G'_0(1)$ \citep{dall,barthr}. 
For the Poissonian degree distribution of RPGs, we can solve Equation \ref{eq:gone} as 
\begin{eqnarray}
S & = & 1 - e^{-\alpha S},
\end{eqnarray}
or, 
\begin{eqnarray}
\alpha S & = & - \ln (1 -S). \label{eq:gsol}
\end{eqnarray}
Figure~\ref{fig:randpercol} shows the solution of Equation~\ref{eq:gsol}. The left panel shows 
that $S = 0$ is the only non-negative solution for $\alpha \leq 1$. 
For $\alpha > 1$, $S$ increases monotonically to the asymptotic value $S = 1$. 
The right panel summarizes the solution of the left panel, showing $S(\alpha)$ vs. $\alpha$. 
\replaced{The percolation threshold, $\alpha_c$, is defined at the transitional point, $\alpha_c = 1$.}{The trainsition of $S(\alpha)$ happens at the percolation threshold for RPGs, $\alpha_c = 1$.}

RGGs also have Poissonian degree distributions. 
The difference from RPGs is that the connections are determined by a connecting hyper-sphere,
depending on spatial dimensionality, while RPGs only depending on the single parameter, $p$.  
\cite{dall} reported the percolation thresholds of RGGs for various dimensions, $d$, 
as $\alpha_c(d=2) = 4.52$, $\alpha_c(d=3) = 2.74$, and $\alpha_c(d=\infty) \simeq 1$
\footnote{Hence, RGGs are equivalent to RPGs in percolation at $d=\infty$.}.

\subsection{Transitivity and 3-point Correlation Function}\label{sec:tri}

\begin{figure}[t]
\centering
\includegraphics[height=1.4 in]{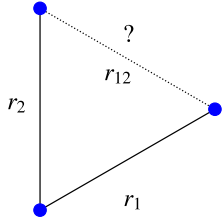}
\caption{The graph schematic for describing the meaning of  {\it transitivity}, $\tau_\Delta$.  
The two connected sides, $r_1$ and $r_2$, form a {\it connected triple}; i.e., a ``$\vee$'' configuration. 
If the other side, $r_{12}$, is also connected, then we refer to this triangular triple as a {\it closed triple}. 
Transitivity is a ratio of closed triples to connected triples. 
This value can be written using $2-$ and $3-$point correlation functions as Equation~\ref{eq:tr}.
}\label{fig:tr}
\end{figure}  

\begin{figure}[t]
\centering
\includegraphics[height=2.8 in]{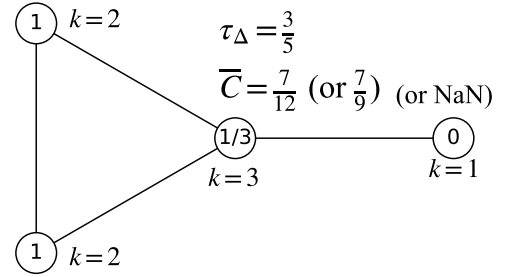}
\caption{The graph schematic for transitivity, $\tau_\Delta$, and average local clustering coefficient, $\overline{C}$. 
The number on each node circle represents the local clustering coefficient, $C_i$, as defined in Equation~\ref{eq:lcc} 
and $k$ the number of neighbors (i.e., {\it degree}).  
The average of local clustering coefficients, $\overline{C}$, is $\frac{7}{12}$ (or, $\frac{7}{9}$ 
when excluding the node with $k=1$ for the average, of which denominator is zero), 
different from the transitivity, $\frac{3}{5}$. 
}\label{fig:slcc}
\end{figure}

Figure~\ref{fig:tr} shows a triple of which two sides, $r_1$ and $r_2$, are connected. This configuration is referred 
to as a {\it connected triple}; if the other side, $r_{12}$, is also connected,  a {\it closed triple}. 
The {\it transitivity}, $\tau_\Delta$, is a triangular density defined using these triple configurations as, 
\begin{eqnarray}
\tau_\Delta & \equiv &  \frac{\textrm{number of closed triples}}{\textrm{number of connected triples}}.    
\end{eqnarray}
For cosmological networks embedded in 3d comoving volume, we can rewrite this equation using correlation functions as,  
\begin{eqnarray}
\tau_\Delta & = &  \frac{\begin{displaystyle} \int_{V_l} d^3r_1 \int_{V_l} d^3r_2~p_3( \bm{r_1}, \bm{r_2}) \Theta(l - r_{12})  \end{displaystyle}}{\begin{displaystyle} \int_{V_l} d^3r_1 \int_{V_l} d^3r_2~p_3( \bm{r_1}, \bm{r_2}) \end{displaystyle}}, \label{eq:tr} \\
p_3( \bm{r_1}, \bm{r_2}) & \equiv & \bar{n}^3 \big[ 1 + \xi(r_1) + \xi(r_2) + \xi(r_{12}) + \zeta(r_1,r_2,r_{12}) \big] , 
\end{eqnarray}
where $\xi(x)$ is 2-point correlation function, $\zeta(x,y,z)$ 3-point correlation function, and $\Theta(x)$ the Heaviside step function (Jeong et al. in prep). 
For RGGs, since their correlation functions vanish, we can derive the transitivities as, 
 \begin{eqnarray}
\tau_\Delta & = & \frac{3}{\sqrt{\pi}} \frac{\Gamma(\frac{d+2}{2})}{\Gamma(\frac{d+1}{2})} \int_0^{\pi/3} \sin^d\theta d\theta, 
\end{eqnarray}
for arbitrary $d$-dimensions \citep[][]{dall}. 

We can define a transitivity-like quantity for each vertex. When assuming that a vertex, $i$, has $k_i$ neighbors and 
the number of triangles centered on this vertex is $\Delta_i$, we can write down a transitivity-like quantity for this vertex, $C_i$, as, 
\begin{eqnarray}
 C_i & \equiv &  \frac{2 \Delta_i}{k_i(k_i-1)}, \label{eq:lcc}
\end{eqnarray}
where $k_i(k_i-1)/2$ is the total number of connected triples (or, ``$\vee$'' configurations) and $\Delta_i$ 
the total number of closed triples on this vertex. This vertex-wise transitivity is referred to as {\it local clustering coefficient} (LCC). 
Then, the {\it average} LCC, $\overline{C}$, can be written as,    
\begin{eqnarray}
 \overline{C} & = & \frac{1}{N} \sum_{i=1}^{N} C_i.
\end{eqnarray}
Due to this averaging process, $\overline{C}$ is biased to the major population of vertices. 
For example, if a galaxy catalog is dominated by field galaxies, the triangular configurations formed by dense group galaxies 
are underrepresented in this statistic, while transitivity is an unbiased network-wise (not, vertex-wise) measurement. 
Figure~\ref{fig:slcc} shows a schema demonstrating the definitions of $\tau_\Delta$ and $\overline{C}$.

\section{Results}

\subsection{Statistics of Graph Configurations}

Figure~\ref{fig:gone} and~\ref{fig:gtwo} show graph statistics of the five Multiverses for the two sample selections\added{:} 
(1) equal mass cut, $M_{\rm cut} = 5 \times 10^{11} h^{-1}$ M$_\odot$, and (2) equal abundance cut, 
$N_{h} = 7,086,717$ as summarized in Table~\ref{tab:sample}. 
\replaced{For each figure, the top-left panel shows giant component fractions ($S_1$), 
the top-right second giant component fraction ($S_2$), the middle-left transitivity ($\tau_\Delta$), 
the middle-right average local clustering coefficient ($\overline{C}$), 
the bottom-left number densities for the connected subcomponents with $s = 2, 3, 4$ ($n_{s=2}$, $n_{s=3}$, $n_{s=4}$), 
and, finally, the bottom-right cumulative number density of all subcomponents with $s \geq 5$ ($n_{s\ge5}$).}{
Each panel shows the giant component fraction ($S_1$; top-left), the second giant component fraction ($S_2$; top-right), the transitivity ($\tau_\Delta$; middle-left), the average local clustering coefficient ($\overline{C}$; middle-right), the number densities for the connected subcomponents with $s = 2, 3, 4$ ($n_{s=2,3,4}$; bottom-left), and the cumulative number density of all subcomponents with $s \geq 5$ ($n_{s \ge 5}$; bottom-right).}

\subsubsection{Equal Mass Cut Sample: $M_{\rm cut} = 5.0 \times 10^{11} h^{-1}$ M$_\odot$}

\begin{figure*}[t]
\centering
\includegraphics[height=6.3 in]{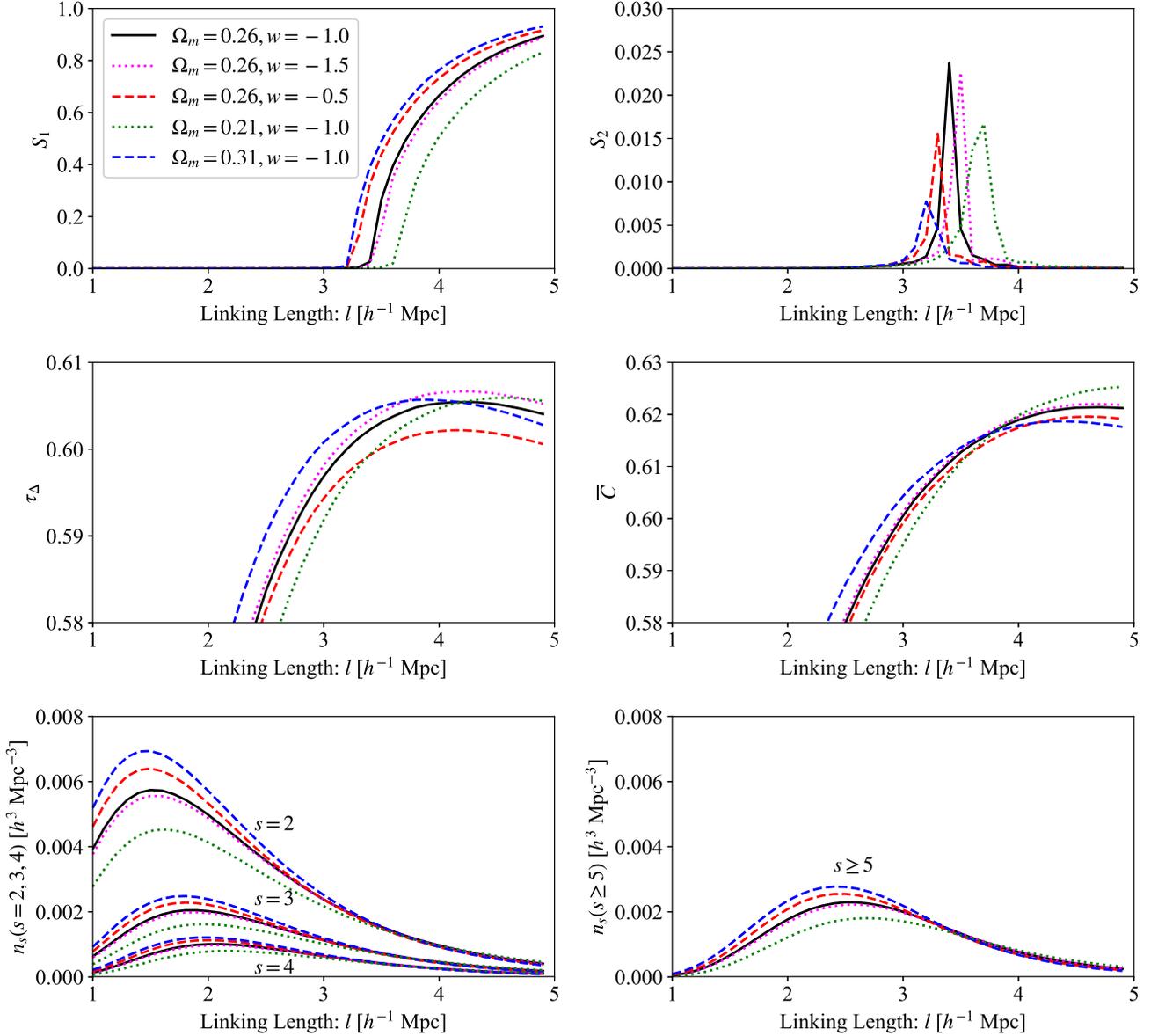}
\caption{The graph statistics vs. the linking lengths for the Multiverse Simulations: Giant Component Fraction ($S_1$, top-left), 
Second Giant Component Fraction ($S_2$, top-right), Transitivity ($\tau_\Delta$, middle-left), 
Average Local Clustering Coefficient ($\overline{C}$, middle-right), 
number densities for the connected subcomponents with $s = 2, 3, 4$ ($n_{s=2}$, $n_{s=3}$, $n_{s=4}$, bottom-left), 
and cumulative number density of all subcomponents with $s \geq 5$ ($n_{s\ge5}$, bottom-right).
As summarized in Table 1, we have two kinds of halo samples, 
using (1) equal mass cut, $M_h \ge 5\times10^{11} h^{-1}$M$_\odot$, 
and (2) eqaul abundance cut, $N_{h} = 7,086,717$. 
This figure is for the equal mass cut sample. 
}\label{fig:gone}
\end{figure*}

For the equal mass cut sample, as shown in Figure~\ref{fig:gone}, 
all graph statistics are quite different enough to discern most of the Multiverses, 
except for the \deleted{model, }DE2\deleted{, } with $\Omega_{\rm m} = 0.26, w = -1.5$ (dotted magenta lines). 
This model shows the least difference among the Multiverse suite from the standard universe 
in two-point statistics and abundances as shown in Figure~\ref{fig:twopoint} and Table~\ref{tab:sample}; 
hence, the most elusive sample to discern statistically. 

The spatial number density directly affects the percolation threshold and comoving densities of connected components. 
More points (vertices) in a fixed volume trivially make the percolation threshold shorter 
since the average distance between point pairs decreases. 
The comoving densities of connected components also increase due to the increment of overall point density.  
Hence, the top and bottom panels in Figure~\ref{fig:gone}, showing the statistics 
of percolation and connected components, are significantly affected by the different abundances.  
When considering most of graph statistics are higher order measurements than the simple one-point statistic, 
any samples without matching abundances are very likely to show trivially different statistics in graph measurements.

\subsubsection{Equal Abundance Sample: $N_{h} = 7,086,717$}\label{sec:eqabun}

\begin{figure*}[t]
\centering
\includegraphics[height=6.3 in]{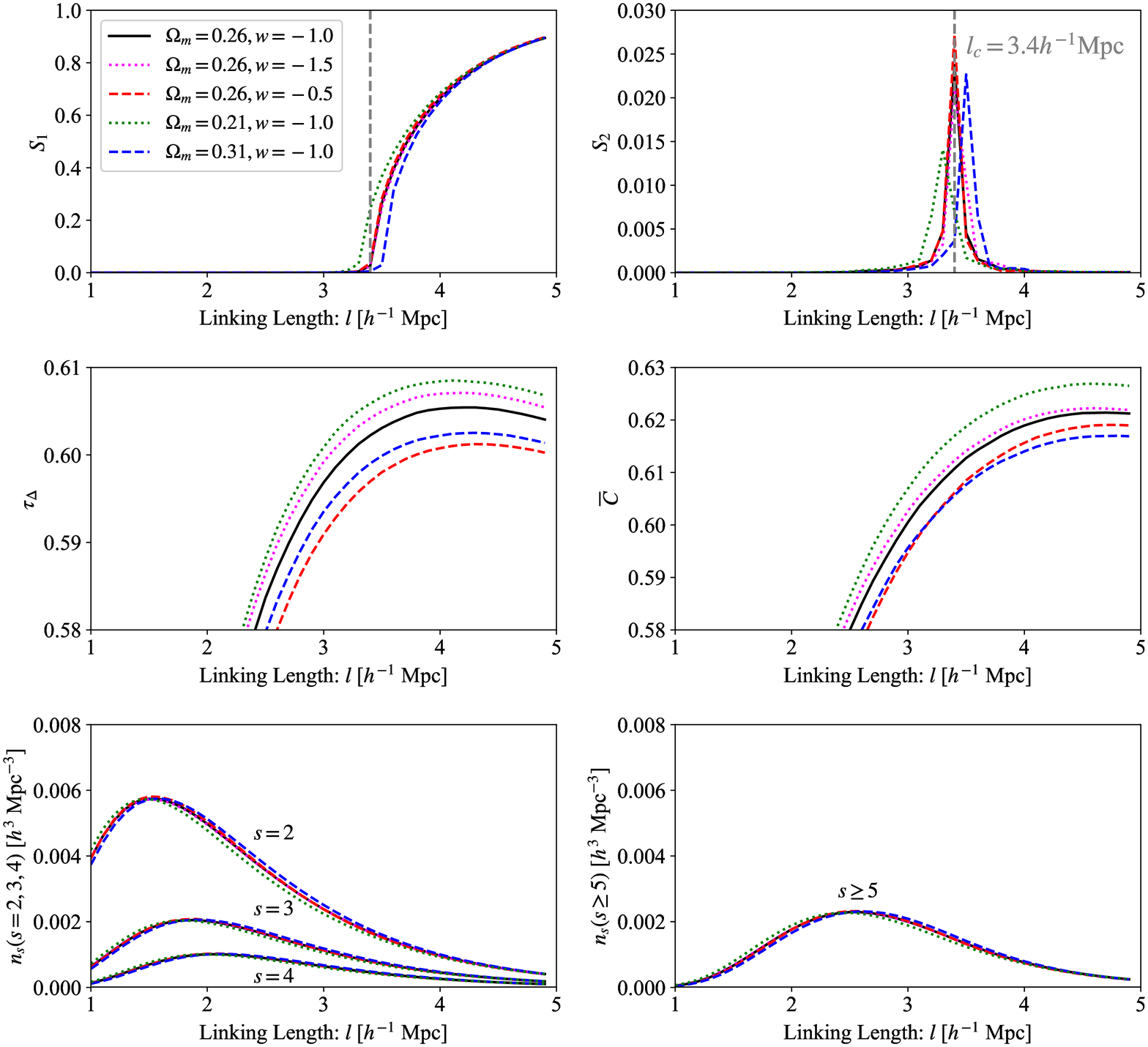}
\caption{The same figure for the equal abundance sample, $N_{h} = 7,086,717$, with Figure~\ref{fig:gone}. 
We can observe that many graph statistics seem degenerate since the abundance effect is removed in this selection. 
}\label{fig:gtwo}
\end{figure*}  

\begin{figure*}[t]
\centering
\includegraphics[height=7.2 in]{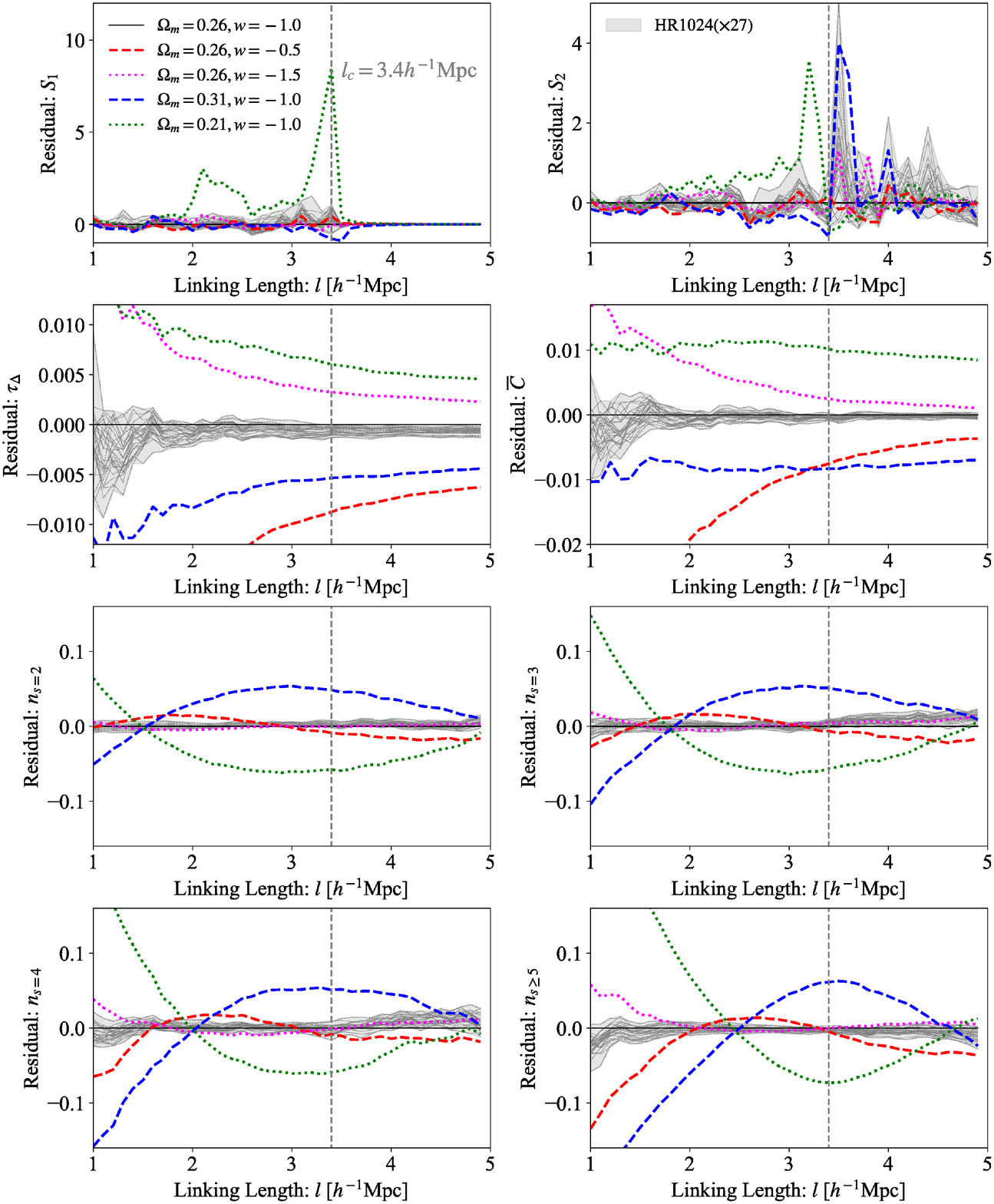}
\caption{The residuals of graph statistics vs. the linking lengths for Figure~\ref{fig:gtwo}. 
From \replaced{Horizon Run}{STD-HR}, we extract 27 \replaced{sub-volumes with the size}{subsamples with the volume} of $1024^3 h^{-3}$ Mpc$^3$. 
The grey shaded area shows the residuals of these 27 subsamples, 
representing the cosmic variances of the standard universe in the Multiverse suite. 
The two statistics, $\tau_\Delta$ at most scales and $n_{s\ge5}$ 
at the percolation threshold (vertical grey dotted line), seem the best discreminants for constraining cosmologies. 
We analyze the graph statistics in details at the percolation threshold in Figure~\ref{fig:gmulti}. 
}\label{fig:gresidual}
\end{figure*}

Figure~\ref{fig:gtwo} shows the graph statistics of equal abundance sample with  
$N_{h} = 7,086,717$, of which comoving density is $n_{\rm h} = 6.6\times10^{-3}  [ h^{-1}$Mpc$]^{-3}$. 
Now, we can observe that many graph statistics seem degenerate 
since the abundance effect is removed in this selection; namely, a good testbed  
how well graph statistics can work as precise discriminators for constraining cosmology. 
 
To better investigate these degenerate-looking features, 
we measure the residuals of graph statistics differing from the standard universe, shown in Figure~\ref{fig:gresidual}. 
We also extract 27 \replaced{sub-volumes with the size}{subsamples with the volume} of $V_{sub} = 1024^3$ $[h^{-1}$Mpc$]^3$ from \replaced{Horizon Run}{STD-HR}
and measure their residuals (grey lines; HR1024) to show the cosmic variances of graph statistics at this size of survey volume.  
The grey shaded area represents the range between the maximum and minimum residuals. 

From the results shown in Figure~\ref{fig:gtwo} and \deleted{Figure~}\ref{fig:gresidual}, we can observe 
that each parameter variation (or, perturbation) of \replaced{Dark Energy}{dark energy}, $w = -0.5, - 1.0, -1.5$, 
and \replaced{Dark Matter}{dark matter} \deleted{(the dominant content for cosmic matter density)}, $\Omega_{\rm m} = 0.21, 0.26, 0.31$, 
affects the graph topology of halo distributions in different ways. 
We describe this in details in the following two separate sections. 

\subsubsection*{Percolation Threshold and Connected Components : \\ Degeneracy in Dark Energy Perturbation}

In Figure~\ref{fig:gtwo}, the top panels show the largest (left, $S_1$) and second largest connected subcomponent (right, $S_2$). 
Interestingly enough, \deleted{all of} the three models with equal $\Omega_{\rm m} = 0.26$\deleted{, but with different dark energy states, 
$w = -0.5, -1.0, -1.5$, DE1, STD, and DE2,} show almost the same percolation curves in $S_1$ and $S_2$ statistics. 
\replaced{Their percolation thresholds}{The percolation thresholds of three models with $\Omega_{\rm m} = 0.26$} are $l_c = 3.4 h^{-1}$Mpc\replaced{. For $\Omega_{\rm m} = 0.21$ and $0.31$, 
their percolation thresholds}{, while those of $\Omega_{\rm m} = 0.21$ and $0.31$} are smaller and larger than $l_c = 3.4 h^{-1}$Mpc\added{, respectively}. 
Hence, for the equal abundance sample, the percolation thresholds seem to only depend on $\Omega_{\rm m}$, 
ignoring the effects of various dark energy states\deleted{, $w = -0.5, -1.0, -1.5$}. 

As a comparison set, we calculate the percolation threshold, $l_c^{RGG}$, for RGG with $d=3$ using its critical threshold value, $\alpha_c = 2.74$, 
\begin{eqnarray}
l_c^{RGG} (d=3) & = &  \Big( \frac{3 \alpha_c}{4 \pi \bar{n} } \Big)^{\frac{1}{3}} \nonumber \\
& = & 4.6 h^{-1} \textrm{Mpc}
\end{eqnarray}
where $\bar{n} = 6.6\times10^{-3}  \big[ h^{-1}$Mpc$\big]^{-3}$. 
Since RGGs have zero correlation functions, the gaps, $| l_c - l_c^{RGG}| = 1.2 h^{-1}$ Mpc, in percolation thresholds 
between RGGs and Multiverse networks are caused by the contributions of all orders of non-zero correlation functions, 
as \cite{zhang18}  have derived using their Probability Cloud Cluster Expansion Theory (PCCET). 
The generating function formulation in Equation~\ref{eq:gone},~\ref{eq:gtwo}, and~\ref{eq:gthree}, 
also show the dependence of percolation threshold on $p_k$ with all orders of  $k$, 
which implicitly reflects the dependence of all correlation functions.

The comoving densities of connected components\replaced{, }{ (}$n_{s=2}$, $n_{s=3}$, $n_{s=4}$, and $n_{s\ge5}$\replaced{,}{)} 
are shown in the bottom panels of Figure~\ref{fig:gtwo}. Their residuals \added{from STD} are plotted in the third and forth rows 
in Figure~\ref{fig:gresidual}. The notable features are the $\cap$ and $\cup$ shapes for DM1 ($\Omega_{\rm m} = 0.31$; \replaced{blue-dashed lines}{blue dashes}) 
and DM2 ($\Omega_{\rm m} = 0.21$; \replaced{green-dotted lines}{green dots}) \deleted{respectively} near the percolation threshold, $l_c = 3.4 h^{-1}$Mpc, in the residual figure. 
In contrast, \replaced{the red and magenta lines, representing DE1 and DE2 with different dark energy parameters, $w = -0.5$ and $ -1.5$,}{DE1 ($w = -0.5$; red lines) and DE2 ($w = -1.5$; magenta lines)}
are marginally separable when considering the cosmic variances (grey area). 
Hence, like the percolation thresholds, the comoving densities of connected components 
\replaced{seem to depend mostly on $\Omega_{\rm m}$, not on $w$.}{depend mainly on $\Omega_{\rm m}$ rather than $w$.}

Finally, the locations of \replaced{intersecting points}{intersection} between the red and magenta lines, 
where the effects of different dark energy parameters\deleted{, $w = -0.5, -1.0, -1.5$,} are nullified in the comoving densities of connected components, 
converge to the percolation threshold, $l_c = 3.4 h^{-1}$Mpc, as the connected component size, $s$, increases. 
For $n_{s\ge5}$, we can observe that the intersecting point is located at the right percolation threshold.  
At this crossing point, the $\cap$ and $\cup$ residual features are, also, most prominent for $n_{s\ge5}$.
The other connected components, $n_{s=2}$, $n_{s=3}$, and $n_{s=4}$, show qualitatively 
the same results with $n_{s\ge5}$. However, their crossing points between the red and magenta lines are located 
with offsets from the percolation threshold and the $\cap$ and $\cup$ residuals are less critical. 
Hence, $n_{s\ge5}$ is the most preferred statistic to represent the properties of connected components as a cosmological discriminator. 

\subsubsection*{Transitivity : Breaking the Degeneracy in Dark Energy Perturbation}
 
The middle panels in Figure~\ref{fig:gtwo} show transitivity ($\tau_\Delta$; left) and local clustering coefficient 
($\overline{C}$; right) for the equal abundance sample. 
Their residuals are plotted in the second row panels in Figure~\ref{fig:gresidual}. 
Unlike the degenerate features of percolation properties, $l_c$ and $n_{s\ge5}$, in the previous section, 
the two triangular statistics, $\tau_\Delta$ and $\overline{C}$, separate all Multiverses quite well. 

As described in \S\ref{sec:tri}, $\overline{C}$ is a biased triangular density, 
while $\tau_\Delta$ an unbiased measurement. 
In addition, the residuals in Figure~\ref{fig:gresidual} are quite consistent for $\tau_\Delta$ in most linking lengths, 
while the residuals of $\overline{C}$ are not. 
Hence, though $\overline{C}$ is a still useful statistic, $\tau_\Delta$ is preferred to $\overline{C}$. 

Overall, Figure~\ref{fig:gresidual} suggests that the two graph statistics, $\tau_\Delta$ and $n_{s\ge5}$, 
measured at the percolation threshold, $l_c = 3.4 h^{-1}$Mpc, are the best statistics to discern different cosmology.

\subsection{ Simple Graph Diagnostics at Big Data Scales : $\alpha$, $\tau_\Delta$, $n_{s\ge5}$ }
 
In the previous section, we have explored the graph properties of Multiverses 
and found that $\tau_\Delta$ and $n_{s\ge5}$ measured at the percolation threshold 
are the best discriminators for constraining different cosmological parameters. 
In this section, we investigate diagnostic diagrams of graph statistics  
and their cosmic variances, depending on survey volume sizes, 
which determine the statistical precision of each diagram. 

\begin{figure*}[t]
\centering
\includegraphics[height=6.2 in]{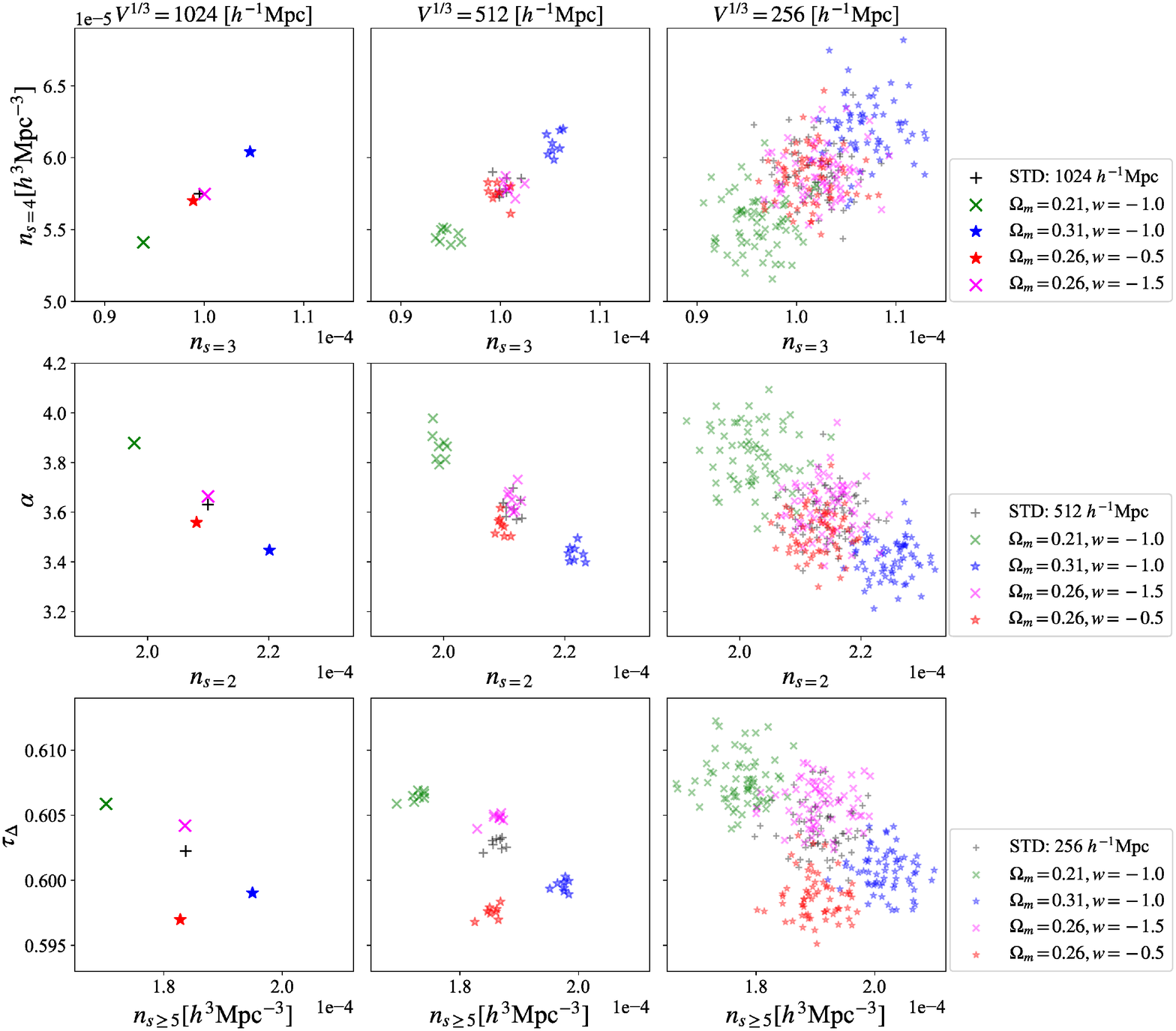}
\caption{The graph diagnostics for the Multiverse samples at the {\it percolation threshold}, $l_c = 3.4 h^{-1}$Mpc. 
We split the total volume, $V^{1/3} = 1024 h^{-1}$ Mpc, into 64 \replaced{sub-volumes with the size}{subsamples with the volume} of $V^{1/3} = 256 h^{-1}$ Mpc (right panels) 
and 8 \replaced{sub-volumes}{subsamples} with $V^{1/3} = 512 h^{-1}$ Mpc (middle panels). 
The single full-volume measurements, $V^{1/3} = 1024 h^{-1}$ Mpc, are shown in the left panels. 
From the implications obtained by these results, we suggest a diagnostic diagram in Figure~\ref{fig:ghr} 
and present its statistical precision using finite-size scaling relations in Figure~\ref{fig:gvar}.   
}\label{fig:gmulti}
\end{figure*}  

Figure~\ref{fig:gmulti} shows three diagnostic diagrams, $n_{s=4}$ vs. $n_{s=3}$ (top), $\alpha$ vs. $n_{s=2}$ (middle), 
and $\tau_\Delta$ vs. $n_{s\ge5}$ (bottom), measured at the percolation threshold\deleted{,} for three different \replaced{volume sizes}{volumes}, 
$V^{1/3} = 256$ (right), $512$ (middle), and $1024$ (left) $h^{-1}$Mpc. 
We split the total volume \deleted{size} of Multiverse simulations, $V^{1/3} = 1024$ $h^{-1}$Mpc, into 64 \replaced{sub-volumes for}{subsamples with} $V^{1/3} = 256$ $h^{-1}$Mpc
and 8 \replaced{sub-volumes for}{subsamples with} $V^{1/3} = 512$ $h^{-1}$Mpc, which show roughly the cosmic variances 
for given \replaced{sub-volume sizes}{subsample volumes} in the diagnostic diagrams. 

We can obtain various implications from the results in Figure~\ref{fig:gmulti}. 
First, the cosmic variance of graph diagnostics for $V^{1/3} = 256$ $h^{-1}$Mpc is too large to properly constrain the cosmological parameters. 
The second-column panels, roughly, suggest that we need a survey volume, $V^{1/3} \ge 512$ $h^{-1}$Mpc. 
Samplings in gigaparsecs scales will be necessary for more precise constraints. 
Therefore, graph analyses for constraining cosmology are inevitably a Big Data science. 
We will present the details about statistical precision of each graph statistic vs. data-size later in a separate paragraph.   
Second, the diagnostic diagrams of $n_{s=4}$ vs. $n_{s=3}$ (top panels) now clearly visualize the degeneracy 
of connected component statistics in dark energy perturbation, elaborately described in \S\ref{sec:eqabun}. 
Using $\alpha$ in the diagnostic diagram of $\alpha$ vs. $n_{s=2}$ (middle panels), 
we have a minor improvement for discerning the different dark energy parameters than the $n_{s=4}$ vs. $n_{s=3}$ diagram, 
but still this diagnostic diagram is not practically useful. 
Finally, as shown in Figure~\ref{fig:gresidual}, the diagnostic diagrams of  $\tau_\Delta$ vs. $n_{s\ge5}$ (bottom panels) 
can separate all of the five Multiverses, though the survey volume \deleted{size} of $V^{1/3} = 256$ $h^{-1}$Mpc 
is still too small to constrain cosmology even in this diagnostic diagram. 
Consequently, including $\alpha$ as a proxy measurement of most commonly used two-point correlation function, 
we suggest a simple set of diagnostics, $\{\alpha, \tau_\Delta, n_{s\ge5}\}$, as a quick look of 
various orders of $n$-points correlation functions for cosmological Big Data sets.

\begin{figure*}[t]
\centering
\includegraphics[height=2.7 in]{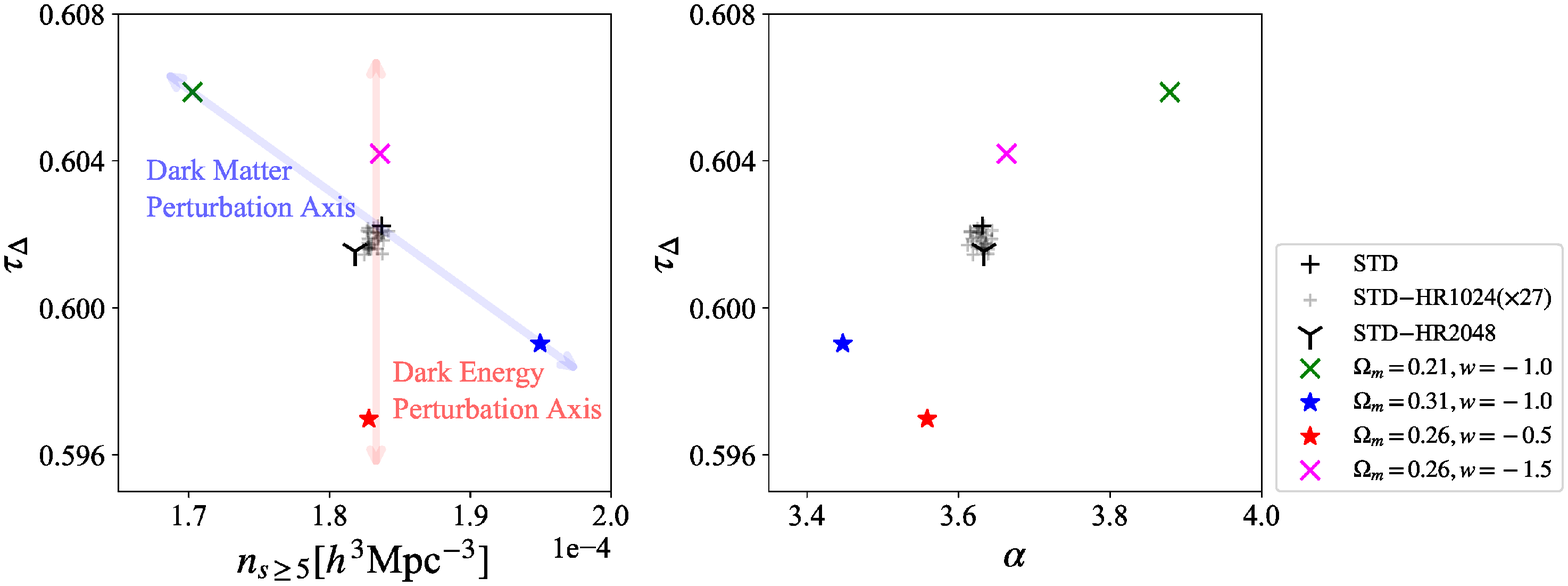}
\caption{The graph diagnostics of $\{ \alpha, \tau_\Delta, n_{s\ge5}\}$ at the {\it percolation threshold}, $l_c = 3.4 h^{-1}$Mpc. 
Like Figure~\ref{fig:gresidual}, we extract 27 \replaced{sub-volumes with the size}{subsamples with the volume} of $1024^3 h^{-3}$ Mpc$^3$ from the Horizon Run data. 
The grey `+' markers, referred to as STD-HR1024, show the graph statistics of these 27 \replaced{sub-volumes}{subsamples}, 
representing the cosmic variances of the diagnostics, which are quite small enough for accurately discerning 
all different Multiverses. The largest sample of STD-HR2048 is composed of 57 millions halos (vertices) with 206 millions connections (edges). 
}\label{fig:ghr}
\end{figure*}

Figure~\ref{fig:ghr} shows our final diagnostic diagrams, representing $\{\alpha, \tau_\Delta, n_{s\ge5}\}$. 
Except for the `Y' marker, all data points are obtained using $V^{1/3} = 1024$ $h^{-1}$Mpc; hence, samplings in a gigaparsec scale. 
The `Y' marker, referred to as STD-HR2048, represents a single selection with $V^{1/3} = 2048$ $h^{-1}$Mpc, 
extracted from \replaced{Horizon Run}{STD-HR}. This largest sample is composed of 57 millions halos (vertices) with 206 millions connections (edges). 
The grey `+' makers, referred to as STD-HR1024$(\times 27)$, represent 27 \replaced{sub-volumes}{subsamples}
with $V^{1/3} = 1024$ $h^{-1}$Mpc, extracted from \replaced{Horizon Run}{STD-HR}, 
showing the cosmic variances of $\{\alpha, \tau_\Delta, n_{s\ge5}\}$ for the standard cosmology 
at the scale of $V^{1/3} = 1024$ $h^{-1}$Mpc. The grey shaded area shown in Figure~\ref{fig:gresidual} is 
equivalent to these grey `+' markers. 

From the diagnostics diagrams in Figure~\ref{fig:ghr}, we can distinguish the most elusive sample, DE2, 
with $\Omega_{\rm m} = 0.26, w = -1.5$ (magenta `x'), from the standard universe (black `+') with a high statistical precision. 
In the $\tau_\Delta$ vs. $n_{s\ge5}$ diagnostics (left panel), the dark energy perturbation moves 
the graph statistics vertically from the standard universe due to the degenerate statistics in percolation and connected components. 
On the other hand, the dark matter, the dominant content for $\Omega_{\rm m}$, perturbation changes all statistics, resulting in 
moving the graph statistics in the oblique axis from the standard universe. 

Since gravity is an all-range force, the variation of $\Omega_{\rm m}$ affects all scales of matter distributions. 
This unique property of gravity changes all graph statistics as shown in many figures through this paper. 
However, since dark energy only expands the space, the effect of dark energy variation should be limited, 
when compared to the effect of gravity. In the graph statistics, this limitation of dark energy is observed 
as the degenerate statistics in percolation and connected components. 
Due to this difference, each parameter perturbation moves the graph statistics along different axis 
as shown in Figure~\ref{fig:ghr}. 

\begin{figure*}[t]
\centering
\includegraphics[height=5.0 in]{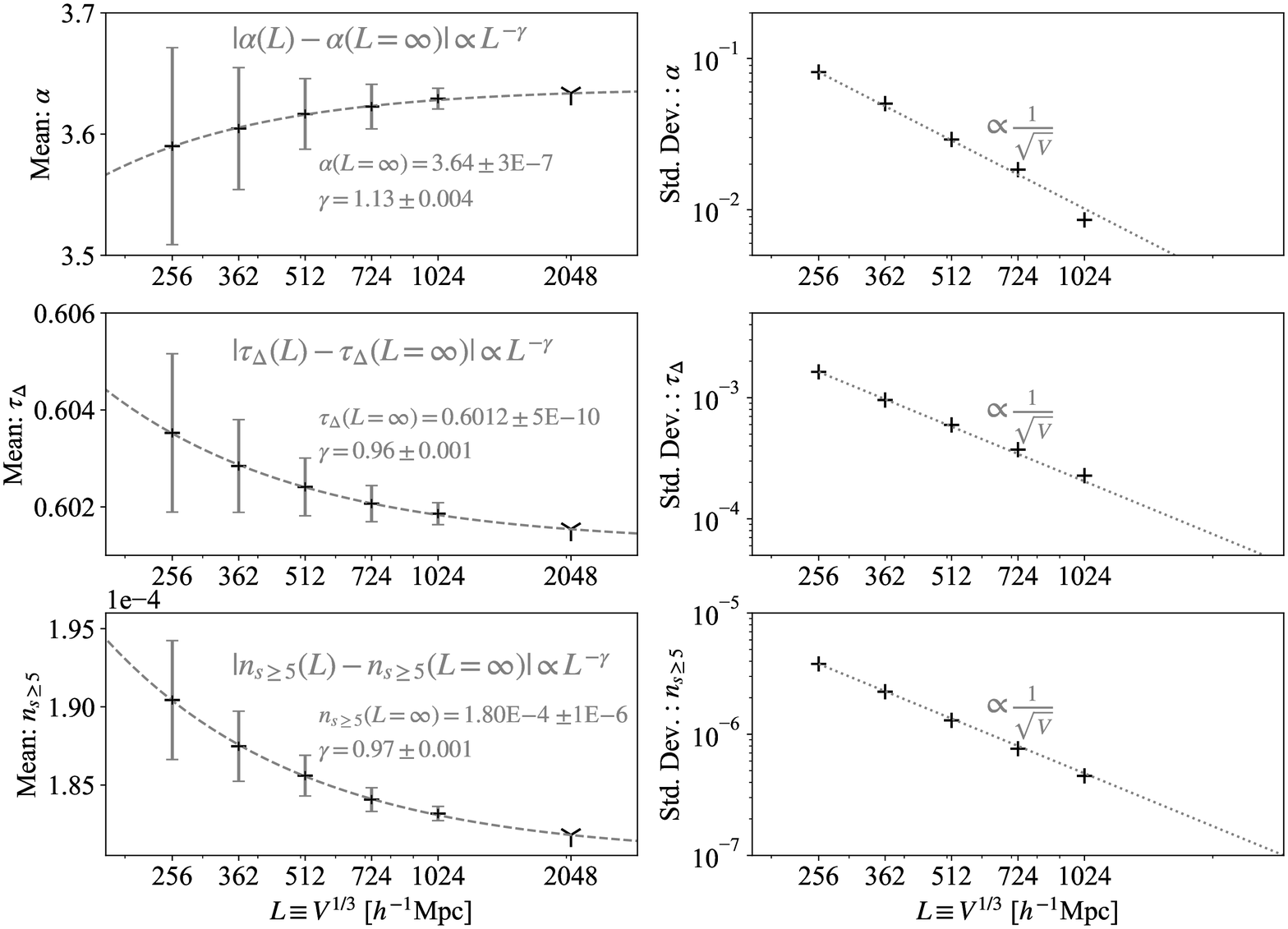}
\caption{The means and standard deviations of $\{ \alpha, \tau_\Delta, n_{s\ge5}\}$ for various volume sizes 
at the {\it percolation threshold}, $l_c = 3.4 h^{-1}$Mpc. 
The numbers of sub-samples for $V^{1/3}$ = 256, 362, 512, 724, and 1024 $h^{-1}$Mpc are $1728$, $512$, $216$, $64$, and $27$ respectively. 
The `Y' marker represents STD-HR2048, also shown in Figure~\ref{fig:ghr}. 
In the left panels, we fit the mean values for each statistic using the finite-size scaling function. 
In the right panels, from the $V^{1/3} = 256 h^{-1}{\rm Mpc}$ results we extrapolate the standard deviation values 
following the scaling relation, $\propto \frac{1}{\sqrt{V}}$ 
(i.e., $\propto L^{-1.5}$), plotted as grey dotted lines. 
Overall, the effects of survey volume sizes are well predictable by finite-size scaling relations with Poissonian variances. 
Notably, this scaling analysis is virtually impossible without modern Big Data tools. 
}\label{fig:gvar}
\end{figure*}

Figure~\ref{fig:gvar} shows how each graph quantity depends on volume sizes for Horizon Run, the largest simulation box. 
The numbers of sub\deleted{-}samples for $L \equiv V^{1/3} $ = 256, 362, 512, 724, and 1024 $h^{-1}$Mpc are 
$1728$, $512$, $216$, $64$, and $27$ respectively. 
In the right panels, we extrapolate the standard deviation values from the results at $V^{1/3} = 256 h^{-1}{\rm Mpc}$, 
following the scaling relation, $\propto \frac{1}{\sqrt{V}}$ (i.e., $\propto L^{-1.5}$; grey dotted lines). 
The measured standard deviations (hence, the cosmic variances; `+' markers) of the graph diagnostics, $\{ \alpha, \tau_\Delta, n_{s\ge5} \}$, 
follow this scaling relation, $\propto \frac{1}{\sqrt{V}}$, quite well. 

For the mean values of $\{ \alpha, \tau_\Delta, n_{s\ge5}\}$, we fit them using the scaling relation, 
\begin{equation}
| \eta(L) - \eta(L = \infty)| \propto L^{-\gamma}\label{eq:fscale}, 
\end{equation}
where $\eta(L)$ is one of $\{ \alpha, \tau_\Delta, n_{s\ge5} \}$ at the system size, $L \equiv V^{1/3} $. 
This scaling relation is also known as finite-size scaling in statistical physics.\footnote{The typical finite-size scaling formula is $| \eta(L) - \eta(L = \infty)|^{-\nu} \propto L$, not Equation~\ref{eq:fscale}; in our scaling convention, $\gamma \equiv \frac{1}{\nu}$.} 
We rewrite Equation~\ref{eq:fscale} in a more practical form as, 
\begin{eqnarray}
\eta(L) =  \epsilon \Big( \frac{L}{L_0} \Big)^{-\gamma}  + \eta_0 - \epsilon, \label{eq:pscale}
\end{eqnarray}
where $\eta_0  = \eta(L_0)$ and $\eta(L = \infty) = \eta_0 - \epsilon$.
The left panels of Figure~\ref{fig:gvar} show the scaling exponents and asymptotic values by using the fitting function, Equation~\ref{eq:pscale}, 
with $L_0 = 2048 h^{-1}{\rm Mpc}$.   
Consequently, the effects of survey volume sizes on the graph diagnostics, $\{ \alpha, \tau_\Delta, n_{s\ge5} \}$,  
are well predictable by finite-size scaling relations with Poissonian variances. 
Notably, this scaling analysis is virtually impossible without modern Big Data tools.

\section{Summary and Discussion}

By utilizing the modern Big Data platform, \textsc{Apache Spark}, 
we have investigated the graph topology of discrete point distributions of dark matter halos for five different universes; 
a suite of Multiverse simulations,  
(1) STD: $\Omega_{\rm m} = 0.26, w = -1.0$, (2) DE1: $\Omega_{\rm m} = 0.26, w = -0.5$, 
(3) DE2: $\Omega_{\rm m} = 0.26, w = -1.5$, (4) DM1: $\Omega_{\rm m} = 0.31, w = -1.0$, 
and (5) DM2: $\Omega_{\rm m} = 0.21, w = -1.0$. 
The equal mass cut sample, selecting halos above $M_{\rm cut} = 5\times 10^{11} h^{-1}$M$_\odot$, 
shows quite different graph statistics, mainly due to their different abundances, 
which affect graph measurements significantly. 
Hence, it is trivial to discern all of the five different Multiverses using graph statistics in this equal mass cut selection. 

The equal abundance sample, selecting halos using $N_{h} = 7,086,717$ 
of which comoving density is $n_{\rm h} = 6.6 \times 10^{-3}$ $[h^{-1}$Mpc$]^{-3}$, 
show degenerate statistics in percolation threshold and connected components for STD, DE1, and DE2. 
This means that the graph statistics related to percolation, 
$\{n_{s=2}$, $n_{s=3}$, $n_{s=4}$, $n_{s\ge5}$, $l_c\}$, 
mostly depend on $\Omega_{\rm m}$, not $w$. 

The degenerate percolation threshold for STD, DE1, and DE2 is $l_c = 3.4 h^{-1}$Mpc, 
different from their corresponding RGG, $l_c^{RGG} = 4.6 h^{-1}$Mpc. 
Since RGG has zero correlation functions, the difference in percolation thresholds, 
$|l_c - l_c^{RGG}| = 1.2 h^{-1}$Mpc, between RGG and Multiverse networks 
is caused by the non-zero correlation functions of all orders.  

This degeneracy can be removed by the triangular statistics, $\tau_\Delta$ and $\overline{C}$.  
Among all graph statistics measured in this paper, $\tau_\Delta$ and $n_{s\ge5}$ 
are the best discriminators for constraining cosmology. 
By including $\alpha$ as a proxy of most commonly used statistic, two-point correlation function, 
we have suggested a graph diagnostics set, $\{\alpha$, $\tau_\Delta$, $n_{s\ge5} \}$, 
as a quick look of various orders of correlation functions at Big Data scales in a computationally cheap way. 
Using the finite-size scalings, we have shown that 
the cosmic means and variances of $\alpha$, $\tau_\Delta$, and $n_{s\ge5}$ are well described 
by various power-laws. 

Future research will investigate the practical observable, {\it galaxies}, at Big Data scales 
since the obvious caveat of this work is the FoF halo catalogs, 
which lack for complex and sophisticated baryonic physics in formation and evolution of galaxies. 
As \cite{hong19} have reported a transitivity anomaly in Lyman alpha emitting galaxies (LAEs), 
implying a strong environmental effect on formation and evolution of LAEs, 
graph statistics of galaxy catalogs are inevitably affected by baryonic physics, 
which could erase the underlying cosmological parameters. 
Hence, we may need to extract more topological features 
from galaxy catalogs  for better constraining cosmology using the state-of-the-art graph analyses. 
Technically, this means that we need to fully utilize both of single machine and 
distributed computing Application Programming Interfaces (APIs).  
The single machine APIs support many feature extractions, but limited to small data sets fit in a single machine, 
while the distributed computing APIs support limited feature extractions, but can handle big data sets.  
Therefore, galaxy catalogs at Big Data scales will be a good challenge 
to fully test the current state-of-the-art graph analyses tools. 

\acknowledgments
Authors acknowledge the Korea Institute for Advanced Study for providing computing resources 
(KIAS Center for Advanced Computation Linux Cluster System).
This work was supported by the Supercomputing Center/Korea Institute of Science 
and Technology Information, with supercomputing resources including technical support 
(KSC-2016-C3-0071) and the simulation data were transferred 
through a high-speed network provided by KREONET/GLORIAD.
\added{SEH was supported by Basic Science Research Program 
through the National Research Foundation of Korea (NRF) 
funded by the Ministry of Education (2018\-R1\-A6\-A1\-A06024977).}

\software{\textsc{Apache Spark} \citep{Zaharia:EECS-2014-12}}

\bibliography{multiverse}

\end{document}